\documentclass{elsart}

\usepackage{hyperref}
\usepackage{epsfig}
\usepackage{graphics}
\usepackage{color}

\begin{document}

\bibliographystyle{unsrt}

\begin{frontmatter}

\title{Deriving the respiratory sinus arrhythmia from the heartbeat time series using Empirical Mode Decomposition}

\author[ifc]{R. Balocchi\corauthref{cor}},
\corauth[cor]{Corresponding author.}
\ead{balocchi@ifc.cnr.it}
\author[ifc]{D. Menicucci},
\author[ifb]{E. Santarcangelo},
\author[ifb]{L. Sebastiani},
\author[ifb]{A. Gemignani},
\author[ifb]{B. Ghelarducci},
\author[ifc]{M. Varanini}

\address[ifc]{Istituto di Fisiologia Clinica, CNR, via Moruzzi 1, 56124 Pisa, Italy}

\address[ifb]{Dipartimento di Fisiologia e Biochimica, Universit\`a di Pisa, via San Zeno 31, 56127 Pisa, Italy}

\begin{abstract}

Heart rate variability (HRV) is a well-known phenomenon whose
characteristics are of great clinical relevance in
pathophysiologic investigations. In particular, respiration is a
powerful modulator of HRV contributing to the oscillations at
highest frequency. Like almost all natural phenomena, HRV is the
result of many nonlinearly interacting processes; therefore any
linear analysis has the potential risk of underestimating, or even
missing, a great amount of information content. Recently the technique of 
Empirical Mode Decomposition (EMD) has been proposed as a new tool
for the analysis of nonlinear and nonstationary data. We applied
EMD analysis to decompose the heartbeat intervals series, derived
from one  electrocardiographic (ECG) signal of 13 subjects, into
their components in order to identify the modes associated with
breathing. After each decomposition the mode showing the highest
frequency and the corresponding respiratory signal were Hilbert
transformed and the instantaneous phases extracted were then
compared. The results obtained indicate a synchronization of order
1:1 between the two series proving the existence of phase and
frequency coupling between the component associated with breathing
and the respiratory signal itself in all subjects.
\end{abstract}

\end{frontmatter}

\section{Introduction}

Oscillations in heart rate on a beat by beat basis are a
well-known phenomenon of great clinical relevance for both
diagnostic and prognostic purposes. Many different factors (i.e.
neural, chemical, hormonal modulations) contribute to the heart
rate variability (HRV) ~\cite{malliani1}. Among them, a major
influence is exerted by the activity of the Sympathetic and
Parasympathetic (vagal) branches of the Autonomic Nervous System
(ANS) ~\cite{taskforce}. In particular, respiration is a powerful
modulator of HRV acting under the control of the ANS
parasympathetic branch; in addition, abnormalities in respiratory
modulation are often an indication of autonomic disfunction
~\cite{respiro,Pinna}.

Traditionally HRV analysis is accomplished, in the frequency
domain, via spectral analysis of the heartbeat time series (RR
intervals) and, in the time domain, by extracting statistical
indexes not related to specific cycle lengths. Parasympathetic
activity is primarily reflected in the high-frequency (HF)
component of the power spectrum (0.15-0.40 Hz) and it is related
to the breathing frequency (respiratory sinus arrhythmia, RSA). More
controversial is the interpretation of the low-frequency (LF)
component (0.04-0.15 Hz), which is considered by some
investigators as a marker of sympathetic modulation
~\cite{Sleight,malliani} and by others as influenced by both the
sympathetic and parasympathetic activity ~\cite{Eckberg}. In the
clinical environment a widely used index of sympathovagal balance
is the ratio LF/HF.

For what concerns time domain indexes, they all measure high
frequency variations in heart rate (and thus they are all highly
correlated). An extensive introduction to HRV measurements can be
found in ~\cite{taskforce,standards}.

Both time-domain and spectral methods share limitations imposed by
the intrinsic nature of the RR series. In fact, changes in
autonomic modulations can occur because of a variety of common
factors such as changing position from sitting to standing,
eating, running, sleeping, wakening. This can induce both a lack
of stationarity in the series, compromising a correct application
of these methods, and a shift of the LF and HF central
frequencies, with a possible superposition of the two bands and a
consequent misleading interpretation.

Other methods estimate RSA through a linear model by using an
adaptive filter scheme having the respiratory signal as reference
input ~\cite{Vara1}. The first limitation of these approaches is
the need of the respiratory signal.  The second concerns the
approximation of the linear model that is not always
acceptable. Like almost all natural phenomena, HRV is the result
of many nonlinearly interacting processes; therefore any linear
analysis has the potential risk of underestimating, or even
missing, a great amount of information content. On the other hand,
complex nonlinear approaches ~\cite{Vara2} do not provide the
tracking capability required in the high non-stationary context of
RR interval series.

Recently the Empirical Mode Decomposition (EMD), a signal
processing technique particularly suitable for nonlinear and
nonstationary series, has been proposed ~\cite{EMD} as a new tool
for data analysis. This technique performs a time adaptive
decomposition of a complex signal into elementary, almost
orthogonal components that do not overlap in frequency. The
extracted components have well behaved Hilbert transforms, from
which the instantaneous frequencies can be calculated.

We applied EMD analysis to decompose the RR series into its
components in order to identify, primarily, the respiratory
oscillation. For comparison purpose, the respiratory signal was
recorded simultaneously with the electrocardiogram (ECG). The existence of phase and
frequency coupling between the RR component associated with
breathing and the respiratory signal itself was then investigated.

\section{Empirical Mode Decomposition}

The empirical mode decomposition is a signal processing technique
proposed in 1998 by Huang  ~\cite{EMD} to extract all the
oscillatory modes embedded in a signal without any requirement of
stationarity or linearity of the data. The goal of this procedure
is to decompose a time series into components with well defined
instantaneous frequency by empirically identifying the physical
time scales intrinsic to the data, that is the time lapse between
successive extrema.

Each characteristic oscillatory mode extracted, named Intrinsic
Mode Function (IMF), satisfies the following properties: an IMF is
symmetric, has a unique \emph{local} frequency, and different IMFs
do not exhibit the same frequency at the same time. In other words
the IMFs are characterized by having the number of extrema and the
number of zero crossings equal (or differing at most by one), and
the mean value between the upper and lower envelope equal to zero
at any point.

The algorithm operates through six steps:
\begin{enumerate}
\item Identification of all the extrema (maxima and minima) of the
series $X$. \item Generation of the upper and lower envelope via
cubic spline interpolation among all the maxima and minima,
respectively. \item Point by point averaging of the two envelopes
to compute a local mean series $m$. \item Subtraction of $m$ from
the data to obtain a IMF candidate $h=X-m$. \item Check the
properties of $h$:
\begin{itemize}
\item if $h$ is not a IMF (i.e. it does not satisfy the previously
defined properties), replace $X$ with $h$ and repeat the procedure
from step 1 \item if $h$ is a IMF, evaluate the residue $r=X-h$.
\end{itemize}
\item Repeat the procedure from step 1 to step 5 by \emph{sifting}
the residual signal. The sifting process ends when the residue $r$
satisfies a predefined stopping criterion.
\end{enumerate}

At the end of the procedure we have a residue $r$ and a collection
of $n$ IMFs, named $h_i$ ($i=1,...,n$). The $h_i$ are generated
being sorted in descending order of frequency and therefore $h_1$
is the one associated with the locally highest frequency.
Furthermore the original $X$ can be exactly reconstructed by a
linear superposition:
$$X = \sum_{i=1}^n h_i + r.$$
Usually to eliminate \emph{riding waves} and obtain symmetric
waves, the sifting process has to be repeated a number of times.
To verify whether $h$ owns the IMF properties (step 5 of the
sifting procedure) we used a range-based criterion: $h$ is
retained as IMF if the range of $m$ is a very low fraction of that
of $h$ $(< 0.001)$.

Analogously, for what concerns the stopping criterion of the EMD
iterations, when all the interesting intrinsic modes of the
original series have been extracted, we checked the range of $r$:
the sifting process ends when the range of the residue is low with
respect to that of the original signal ($<10\%$).

\begin{figure}
\begin{center}
\includegraphics[angle=0,width =0.8\columnwidth]{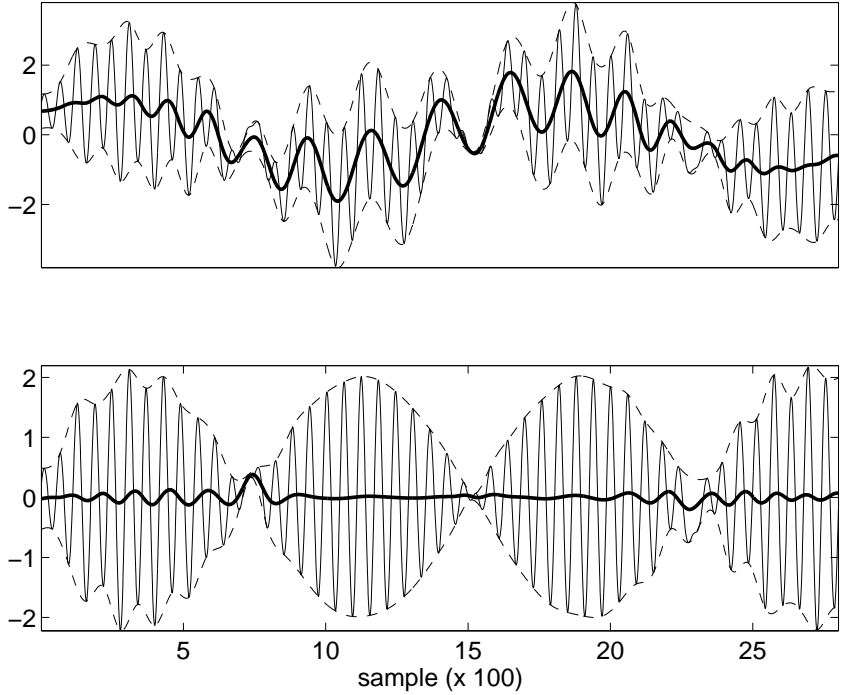}
\caption{Upper panel: the original signal with upper and lower
envelope. The thick line represents the point by point mean value
of the envelopes. Lower panel: the signal $h$ after few
iterations. The iterations continue until $h$ becomes a IMF.}
\label{ES1}
\end{center}
\end{figure}

\begin{figure}
\begin{center}
\includegraphics[angle=0,width =0.8\columnwidth]{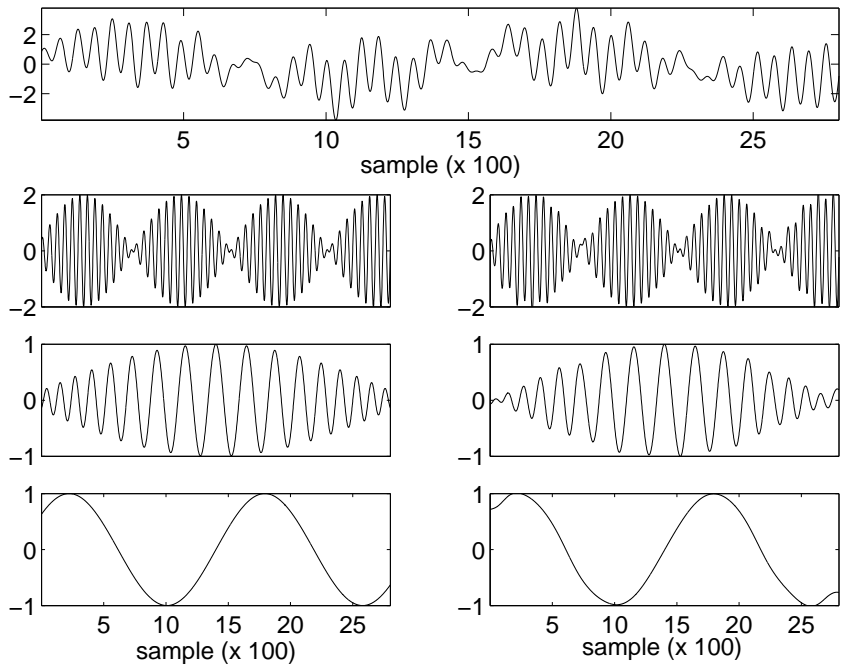}
\caption{Upper panel: the artificial series of 2800 points
obtained by linear combination of three oscillatory signals shown
in the left column. Right column: the IMF extracted using EMD.}
\label{ES2}
\end{center}
\end{figure}

Figure ~\ref{ES1} shows the starting point of a signal
decomposition and the IMF candidate obtained after few iterations.
Figure ~\ref{ES2} shows an example of EMD decomposition of a
simulated series obtained by linear composition of three different
nonstationary oscillations.

The EMD procedure, according to the above specifications, was used
for the decomposition of the RR interval series (tachograms) as
described in the next section.

\section{Experimental procedure}
Thirteen healthy volunteers underwent an experimental session in
which one electrocardiographic (ECG) derivation and the
respiratory signal were simultaneously and continuously acquired
at 1000 Hz sampling frequency. The ECG was recorded using standard
electrodes, while the respiratory signal was detected through a
polymeric piezoelectric dc-coupled transducer inserted into a belt
wrapped around the chest. During the experimental session the
subjects were comfortably sitting and breathing freely.

From the ECG signal, R waves, indicating the ventricular
depolarization, were detected to obtain the tachogram, that is the
series of the time intervals between the occurrence of two
consecutive R peaks. The tachogram was then interpolated at 1 Hz,
to obtain uniformly time spaced data.

The respiratory signals were bandpass filtered in the respiratory
range 0.15-0.4 Hz ~\cite{taskforce} to remove artifacts mainly due
to the positioning of the chest belt and its mechanical
properties. The respiratory signal was then decimated according to
the corresponding interpolated tachogram, thus making the two
series syncronous. The tachograms of the 13 subjects were
decomposed using EMD and each first component was analyzed
together with the corresponding respiratory signal.

\section{Instantaneous frequency analysis}

The local frequency of each IMF can be expressed by the
instantaneous frequency defined by virtue of the analytic signal
of the original data.

Given a real function $f(t)$ with Fourier transform $F(\nu)$, the
analytic signal $f_a(t)$ is defined as the inverse Fourier
transform of the positive frequency part of $F(\nu)$:

$$f_a(t)=2\,\int_0^{+\infty} F(\nu) e^{j\,2 \pi \nu t} dv.$$
The real part of $f_a(t)$ is just $f(t)$ while the imaginary part
defines the Hilbert transform of $f(t)$ ($f_H(t)$). Thus $f_a(t) =
f(t)+ j f_H(t) = a(t)\exp{j \phi(t)}$
in which:

$$a(t) = (f^2(t) + f_H^2(t))^\frac{1}{2} \quad \quad \phi(t) = \arctan \frac{f_H
(t)}{f(t)}.$$

The instantaneous frequency is defined as the rate of phase change:

$$\nu(t) = \frac{d\phi(t)}{dt}.$$

We investigated the phase relationship between the first IMF and
the respiratory signal by computing the instantaneous phases
$\phi_1$ and $\phi_2$ from both signals and  looking at their
behavior by plotting $\phi_1$ mod $2\pi$ versus $\phi_2$, mod
$2\pi$ (bi-plot).

When a phase locking exists, the points in the bi-plot are not
uniformly distributed and some structure is visible
~\cite{Palus,synchro,Toledo,synchro1}. In case of linear phase
coupling, the points in the bi-plot are ideally aligned along a
straight line, in practice they cluster along  a \emph{stripe}
whose slope indicates the frequency ratio (synchronization order)
of the two signals.

The \emph{stripe} slope was  estimated through a generalized
regression, that is detecting the direction of the eigenvector
associated to the highest eigenvalue (maximum variance direction)
via a Principal Components Analysis (PCA) procedure
~\cite{multivariata}. The goodness of the relationship was
measured by the residual variance, an index of the \emph{spread}
of points along the orthogonal direction. The instantaneous
frequencies were computed using the toolbox \emph{Time-frequency
toolbox for Matlab} (TFTB) ~\cite{TFTB}.

\section{Results}

The tachograms of all the 13 subjects were decomposed using EMD.
The number of the oscillatory modes extracted ranged between 5 and
8 (median~7).

\begin{figure}
\begin{center}
\includegraphics[angle=0,width =0.8\columnwidth]{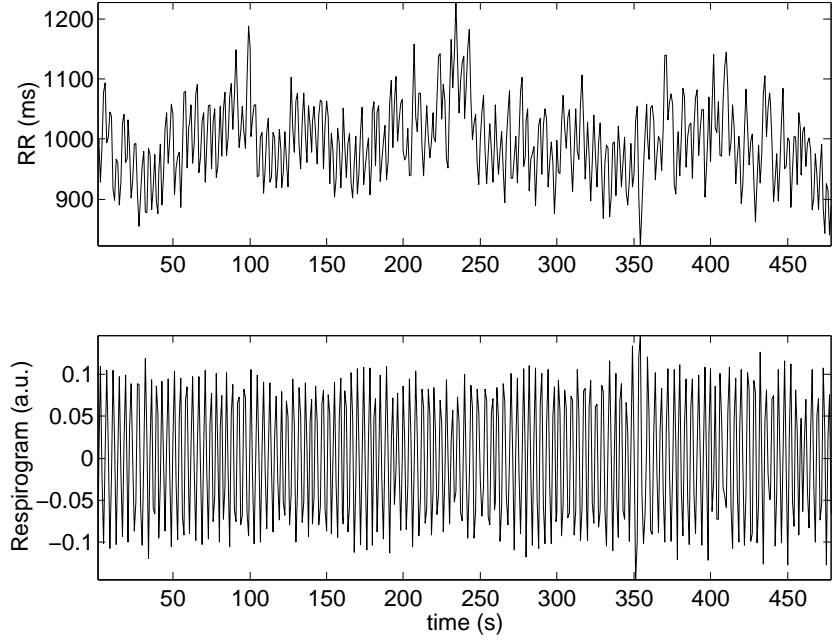}
\caption{An example of tachogram (upper) with simultaneous respiratory time series (lower).} \label{RR_Resp}
\end{center}
\end{figure}

Since the breathing frequency is the highest frequency
contributing to HRV, the first IMF (IMF1) generated was used for
comparison with the recorded respiratory signal. For each subject
IMF1 and the recorded respiratory signal were Hilbert-transformed
and the instantaneous phases computed. Figure ~\ref{RR_Resp} shows
the tachogram and and the simultaneous respiratory signal of one subject.

\begin{figure}
\begin{center}
\includegraphics[angle=0,width =0.8\columnwidth]{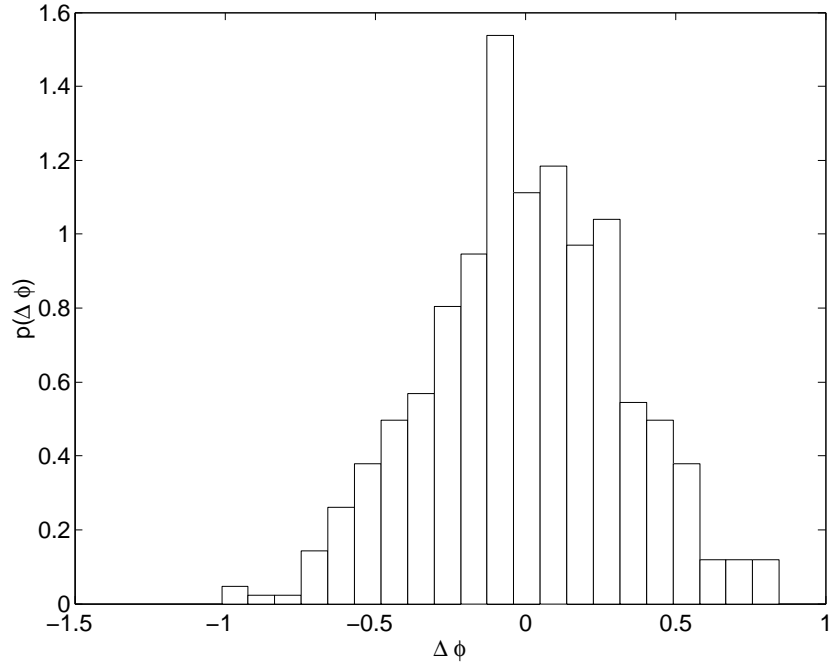}
\caption{Distribution of phase differences between recorded respirogram and IMF1.} \label{phist}
\end{center}
\end{figure}

\begin{figure}
\begin{center}
\includegraphics[angle=0,width =0.8\columnwidth]{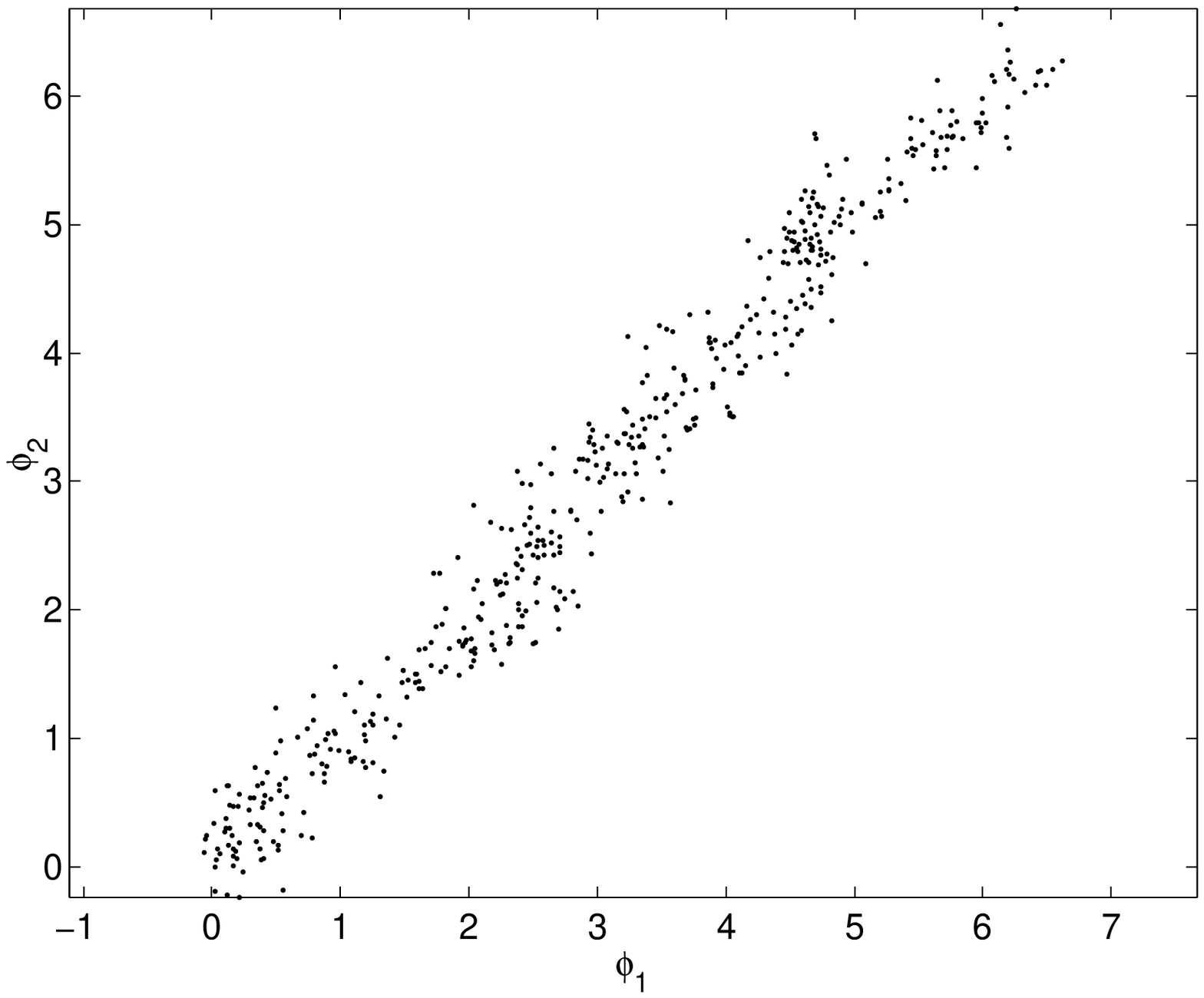}
\caption{Relationship between respirogram anda IMF1 phases.} \label{p1p2}
\end{center}
\end{figure}

The relative phase distribution $\Delta\phi$ = $\phi_1 - \phi_2$ shown in figure ~\ref{phist} exhibits a sharp peak suggestive of a coupling between $\phi_1$ and $\phi_2$. An example of the close correspondence between the two series of instantaneous frequencies is presented in figure ~\ref{p1p2}.

The average frequency ratio between each couple of signals is
$1.015 \pm 0.017$ which is very close to 1, as expected for a
synchronization of order 1:1. The average residual variance is
3.7\%, indicating a good linear relationship.

\section{Discussion}

The Empirical Mode Decomposition technique allows the analysis of
complex, nonlinear and nonstationary time series, through their
decomposition into a limited number of \emph{elementary} modes
having interesting local properties. In the EMD applications
reported in literature ~\cite{waves,Gastrico,Neto,Pressioni}, the
extracted modes are speculatively associated with  specific
physical or physiological aspects of the phenomenon investigated.
In this application we demonstrated the association of the first
IMF extracted from a tachogram with the simultaneously recorded
respiratory signal.

The EMD implementation required some computational skills
essentially to cope with the selection of the stopping criteria
for both sifting and EMD processes and the management of the end
points for cubic splines interpolation. The selected criteria were
not necessarily unique: in any case the procedure is robust with 
respect to a variety of different solutions.

\begin{figure}
\begin{center}
\includegraphics[angle=0,width =0.8\columnwidth]{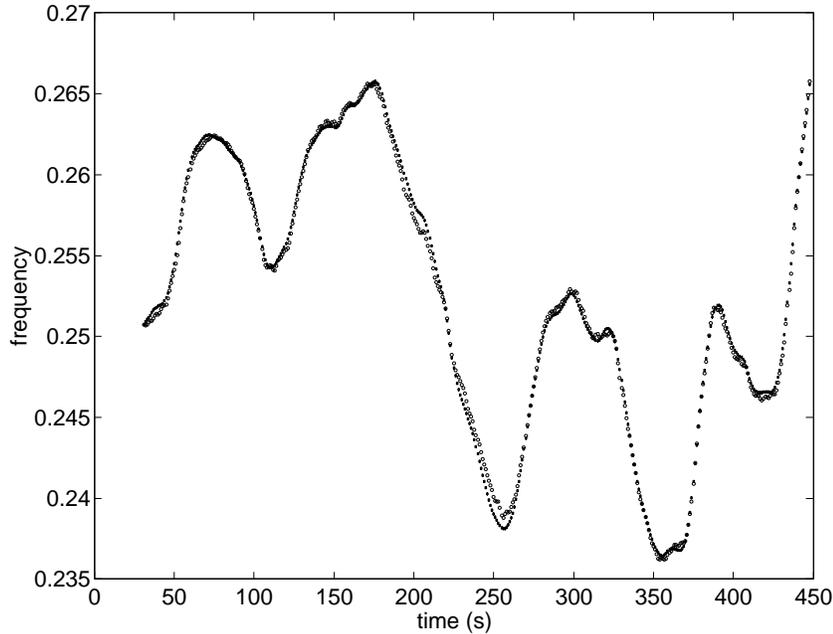}
\caption{Instantaneus frequencies obtained by Hilbert transform of
respirogram and IMF1.} \label{f1f2}
\end{center}
\end{figure}

The peculiar property of each IMF of having a single local
frequency was particularly suitable to compute the instantaneous
phases and frequencies after Hilbert transformation. According to
~\cite{Palus}, in order to verify the synchronization between the
two series, we focused on the phases relationship. We preferred to
work with instantaneous phases instead of frequencies for that the
former resulted to be more stable in time. The series of
frequencies shown in figure ~\ref{f1f2} are indeed not
instantaneously derived, but computed with a smoothing factor from
the toolbox TFTB.

The linear relationship and the generalized regression slope
closest to 1 confirmed the synchronization between $\phi_1$ and
$\phi_2$ and therefore the substantial correspondence between the
IMF1 and respiratory time series. The importance of this finding
is essentially two-fold for the physiologist.  First, the
instantaneous respiratory contribution can now be available even
in all those retrospective studies where the breathing signal was
lacking. Second, the possibility of selectively and efficiently
eliminate the respiratory component from the tachogram makes
possible a more accurate evaluation of the LF role in ANS studies.

Our goal is now to proceed with this approach in the attempt of
attributing a physiological meaning to the other modes embedded in
the tachogram in order to improve the knowledge of the
pathophysiological mechanisms underlying heart rate variability.

\end{document}